\documentclass[aps,preprint]{revtex4}

\usepackage[dvips]{graphicx}

\newcommand{\ch}{{\cal H}}
\newcommand{\ce}{{\cal E}}
\newcommand{\cpt}{\cal PT}

\begin{document}

\date{\today}

\title{Open quantum systems with loss and gain}
\author{Hichem Eleuch$^{1}$\footnote{email: heleuch@physics.mcgill.ca} and 
Ingrid Rotter$^{2}$\footnote{email: rotter@pks.mpg.de}}

\address{
$^1$Department of Physics, McGill University, Montreal, Canada H3A 2T8}
\address{
$^2$Max Planck Institute for the Physics of Complex Systems,
D-01187 Dresden, Germany }

\begin{abstract}
We consider different properties of small open quantum systems coupled
to an environment and
described by a non-Hermitian Hamilton operator.  Of special interest
is the non-analytical behavior of the eigenvalues in the vicinity of
singular points, the so-called exceptional points (EPs), at which the
eigenvalues of two states coalesce and the corresponding 
eigenfunctions are linearly
dependent from one another. The phases of the eigenfunctions are not
rigid in approaching an EP and providing therewith the possibility to 
put information from the environment into the system.  
All characteristic properties 
of non-Hermitian quantum systems hold true not only
for  natural open quantum systems that suffer {\it loss} due to their
embedding into the continuum of scattering wavefunctions.  They appear
also in systems coupled to different layers some of which provide 
{\it  gain} to the system. Thereby  gain and loss, respectively, 
may be fixed inside every layer, i.e. characteristic of it.   

\end{abstract}

\pacs{\bf }
\maketitle

\section{Introduction}
\label{intr}

During last about 15 years, quantum systems described by a  non-Hermitian 
Hamilton operator with ${\cal PT}$ symmetry and entirely real
eigenvalues in a broad
parameter range are considered intensely in literature, e.g. 
\cite{bender1,bender2,specjpa}. According to theory,  
the eigenvalues of the non-Hermitian Hamiltonian become complex if 
${\cal PT}$ symmetry is broken.
Using the formal equivalence of the quantum mechanical Schr\"odinger
equation to the optical wave equation in 
${\cal PT}$-symmetric optical lattices
\cite{equivalence1,equivalence2,equivalence3,equivalence4}, 
this theoretical result has been tested successfully 
about 5 years ago \cite{mumbai1,mumbai2}, see also \cite{kottos1}. 

Already in 2010 the relation between ${\cal PT}$-symmetry breaking 
and the existence of exceptional points (EPs) in quantum systems, 
that are described
by a non-Hermitian operator, is discussed  \cite{jopt}. 
The EP is, according to the definition by Kato \cite{kato}, 
a singular point. In an open quantum system, it appears   
in the continuum of scattering wavefunctions 
into which the considered (local) system is embedded.
Here, two eigenvalues of the non-Hermitian Hamiltonian coalesce and
the corresponding two eigenfunctions are linearly dependent from one
another, see e.g. the review \cite{top}. As a consequence, the
eigenvalues show a  non-analytical behavior 
in the neighborhood of an EP and the two eigenfunctions are mixed
strongly (entangled) in a finite parameter range around an EP. 
The role of EPs for $\cpt$-symmetry breaking has been discussed also
in other papers, see \cite{specjpa}.

An open quantum system is described, in a very natural manner, by a 
non-Hermitian Hamilton operator $\ch$ the complex eigenvalues $\ce_k = E_k +
\frac{i}{2} \Gamma_k$  of which
provide not only the energies $E_k$ of the states but also their lifetimes
which are inverse proportional to the widths $\Gamma_k$. The
eigenvalues and eigenfunctions contain the feedback from the
environment of scattering wavefunctions onto the system properties. 
At low level density, the feedback can be neglected to a good
approximation, and the numerical results obtained by using  $\ch$ 
agree well with those of the standard Hermitian quantum physics.      
At high level density, however, the feedback cannot be neglected and
the eigenvalues and eigenfunctions of  $\ch$ may differ dramatically
from those obtained from a Hermitian Hamiltonian.  The results being
often counterintuitive, appear not only in theoretical studies. Quite
the contrary, mostly they are obtained initially in experimental
studies. They are shown to be 
caused by EPs, see e.g. the review \cite{top} and references therein.    
The results of further calculations on the basis of a schematical
model show  that an EP influences not only the eigenvalues and
eigenfunctions of a non-Hermitian Hamiltonian in a 
finite parameter range around its position but is itself also influenced by
another nearby state. As a consequence, the system achieves a
dynamical phase transition, i.e.
the eigenstates of $\ch$ lose their spectral relation to the original states 
of the system (at low level density) \cite{elro4}.    

The question arises therefore whether or not the results observed
experimentally in \cite{mumbai1,mumbai2,kottos1} are 
characteristic of $\cpt$ symmetry and its breaking or are they  solely
a property related to the non-Hermiticity of the Hamiltonian.
In the present paper we try to find an answer to this question.
To begin with, we provide in Sect. \ref{ham}
the non-Hermitian Hamiltonian $\ch^{(2)}$ of an open quantum system,
by restricting to altogether two states,  and discuss its eigenvalues 
and eigenfunctions in the following Sects. \ref{eigenv} and
\ref{eigenf}, respectively.
In Sect. \ref{crosssection} the question is considered how much
information can be extracted from a study of the cross section ($S$-matrix). 
In the following Sect. \ref{special}, the meaning of an imaginary 
coupling term between the states of an open system and its
environment is discussed  and, eventually, 
gain is included in the description of an open quantum system in
Sect. \ref{gain}. Some conclusions are drawn in the last section.

\section{Hamiltonian of the natural open quantum system}
\label{ham}

In an open quantum system, the discrete states  described by a
Hermitian Hamiltonian $H^B$, are embedded into the continuum of scattering
wavefunctions, which exists always and can not be deleted. 
Due to this fact the discrete states turn into
resonance states the lifetime of which is usually finite. 
The Hamiltonian $\ch$ of the open quantum system reads \cite{top}
\begin{eqnarray}
\label{ham1}
\ch & = & H^B + V_{BC} G_C^{(+)} V_{CB} 
\end{eqnarray}
where $V_{BC}$ and $V_{CB}$ stand for the interaction between system
and environment and $  G_C^{(+)} $ is the Green function in the
environment. The so-called internal (first-order) interaction 
between two states $i$ and $j$ is
involved in $H^B$ while their external (second-order) interaction via the
common environment is described by the last term of (\ref{ham1}). 
The eigenvalues of $\ch$ are complex and
provide not only the energies  of the states but also their lifetimes
(being inverse proportional to the widths).
  
Generally, the coupling matrix elements of the external interaction consist of 
the principal value integral    
\begin{eqnarray}
{\rm Re}\; 
\langle \Phi_i^{B} | \ch |  \Phi_j^{B} \rangle 
 -  E_i^B \delta_{ij} =\frac{1}{2\pi} 
 {\cal P} \int_{\epsilon_c}^{\epsilon_{c}'} 
 {\rm d} E' \;  
\frac{\gamma_{ic}^0 \gamma_{jc}^0}{E-E'} 
\label{form11}
\end{eqnarray}
which is real, and the residuum
\begin{eqnarray}
{\rm Im}\; \langle \Phi_i^{B} | \ch |
  \Phi_j^{B} \rangle =
- \frac{1}{2}\; 
 \gamma_{ic}^0 \gamma_{jc}^0 
\label{form12}
\end{eqnarray}
which is imaginary \cite{top}. Here, the $\Phi_i^{B}$ and  $E_i^B$ are the
eigenfunctions and (discrete) eigenvalues, respectively, of
the Hermitian Hamiltonian $H^B$ which describes the states in the subspace of
discrete states without any coupling to the environment. The
$\gamma_{i c}^0  \equiv
\sqrt{2\pi}\, \langle \Phi_i^B| V | \xi^{E}_{c}
\rangle $
are the (energy-dependent) coupling matrix elements  
between the discrete states $i$ of the system and the environment of
scattering wavefunctions $\xi_c^E$. The $\gamma_{i c}^0$ have to be
calculated for every state $i$ and for each channel $c$ 
(for details see \cite{top}). 
When $i=j$, (\ref{form11}) and (\ref{form12}) give the selfenergy of
the state $i$. 
The coupling matrix elements (\ref{form11}) and (\ref{form12})
(by adding $E_i^B \delta_{ij}$ in the first case) 
are often simulated by complex values  $\omega_{ij}$.

\section{Eigenvalues of the non-Hermitian Hamiltonian} 
\label{eigenv}

In order to study the interaction of two states via the common environment it
is convenient to start from two resonance states (instead of two
discrete states). 
Let us consider, as an example, the symmetric $2\times 2$ matrix 
\begin{eqnarray}
{\cal H}^{(2)} = 
\left( \begin{array}{cc}
\varepsilon_{1} \equiv e_1 + \frac{i}{2} \gamma_1  & ~~~~\omega_{12}   \\
\omega_{21} & ~~~~\varepsilon_{2} \equiv e_2 + \frac{i}{2} \gamma_2   \\
\end{array} \right) 
\label{form1}
\end{eqnarray}
the diagonal elements of which are the two complex eigenvalues 
$ \varepsilon_{i}~(i=1,2)$ of a non-Hermitian operator ${\cal H}^0$
(with $\gamma_i \le 0$ for the decay width of the state $i$ of the 
open quantum system). The $e_i$ and  $\gamma_i$ denote the 
energies and widths, respectively, of the two states when their
interaction via the continuum vanishes, $\omega_{ij} =0$. 
(Note that the width $\gamma_i$ has the dimension of energy
$E$ while the dimension of the coupling matrix elements  
$\gamma_{i c}^0$ defined in (\ref{form11}) and (\ref{form12})
is $\sqrt{E}$ according to the definitions used usually in literature).
The $\omega_{12}=\omega_{21}\equiv \omega$ stand for
the coupling of the two states via the common environment. The
selfenergy of the states is assumed to be included into the $\varepsilon_i$.

The two eigenvalues of ${\cal H}^{(2)}$ are
\begin{eqnarray}
\ce_{i,j} \equiv E_{i,j} + \frac{i}{2} \Gamma_{i,j} = 
 \frac{\varepsilon_1 + \varepsilon_2}{2} \pm Z ~; \quad \quad
Z \equiv \frac{1}{2} \sqrt{(\varepsilon_1 - \varepsilon_2)^2 + 4 \omega^2}
\label{int6}
\end{eqnarray}
where   $E_i$ and $\Gamma_i \le 0$ stand for the
energy and width, respectively, of the eigenstate $i$. 
Resonance states with nonvanishing widths $\Gamma_i$ 
repel each other in energy  according to the value of Re$(Z) $
while the widths bifurcate according to the value of Im$(Z)  $.
The two states cross when $Z=0$. This crossing point is 
a singular point called mostly
exceptional point (EP) according to the definition of Kato \cite{kato}.
Here, the two eigenvalues  coalesce, $\ce_{1}=\ce_{2}$, and the S
matrix has a double pole \cite{marost1,marost2,naz1,naz2}. 
Another notation for the
EP is {\it branch point in the complex plane}
\cite{marost1,marost2,seba1,seba2,naz1,naz2}.  
In the neighborhood of an EP, the parameter
dependence of the eigenvalues $\ce_i$ is non-analytical, especially
that of the widths $\Gamma_i$ (for illustration see Fig. 4 in \cite{seba3}). 
That means, Fermi's golden rule is violated under the influence of an EP.

The condition $Z=0$ cannot be fulfilled for any two discrete eigenvalues 
(with $e_1\ne e_2$) of the Hermitian operator $H^B$.  
This result is known in standard Hermitian quantum physics since many
years: two discrete states  avoid always crossing \cite{landau1,landau2}.  
At the critical parameter value of the avoided crossing,
a geometrical phase, the Berry phase \cite{berry1,berry2}, appears.

In difference to this, the condition  $Z=0$ can generally be fulfilled for 
resonance states (when their widths $\gamma_i$ are sufficiently large). 
This fact causes, on the one hand, differences between open and closed 
systems. Among others, the geometrical phase related to an EP
in the two-level case, differs from the Berry phase by a factor 2,
e.g. \cite{top}, such that the topology of an open system differs 
from that of a closed system. 
On the other hand, discrete states and resonance states show 
an analog behavior related to  crossing: discrete and narrow
resonance states avoid crossing while
broad resonances avoid overlapping. At high level
density where one naively would expect a strong overlapping, the
resonances avoid each other: one resonance accumulates almost the
whole sum of the widths of all resonances while the remaining ones
become nearly stable as illustrated in \cite{cassing}  for altogether
four resonances by means of the corresponding S-matrix poles.
This example is an illustration of the fact that,
in difference to discrete states, the resonances can avoid each
other not only by means of level repulsion caused by Re$(Z)$ but also by width
bifurcation caused by  Im$(Z)$. It is  possible that the states
cross in energy according to  Re$(Z)=0$ while their widths bifurcate
according to  Im$(Z)\ne 0$ (for details see \cite{top}). 
Due to width bifurcation, even bound states in
the continuum (with vanishing width $\Gamma_i$) may appear,
see e.g. \cite{marost1,marost2,bicrosa} and references therein. 

An additional remark should be added here: 
the condition $Z=0$  is fulfilled at only one point in the continuum, 
the EP, at the most. It is therefore of measure zero. However, 
avoided resonance overlapping caused by it
in its neighborhood determines the dynamics of open quantum systems,
see \cite{top}.

\section{Eigenfunctions of the non-Hermitian Hamiltonian}
\label{eigenf}

As discussed in the foregoing section \ref{eigenv},
an EP is  defined by the coalescence of two eigenvalues, $\ce_1 =
\ce_2$,  of the non-Hermitian operator $\ch^{(2)}$. The corresponding 
eigenfunctions $\Phi_{1,2}$ of ${\cal H}^{(2)}$
of the two crossing states are linearly dependent from one another at
the EP, 
\begin{eqnarray}
\Phi_1^{\rm cr} \to ~\pm ~i~\Phi_2^{\rm cr} \; ;
\quad \qquad \Phi_2^{\rm cr} \to
~\mp ~i~\Phi_1^{\rm cr}   
\label{eif5}
\end{eqnarray}  
according to analytical  as well as numerical and experimental
studies  (for details and examples see \cite{top,fdp1,berggren}).
The eigenfunctions $\Phi_k$ of the non-Hermitian $\ch^{(2)}$ 
are biorthogonal because $ H
|\Phi_k\rangle = \ce_k |\Phi_k\rangle$ and  $\langle \Phi_k^* | H = 
\ce_k  \langle \Phi_k^* |$ (where
$\ce_k$ is an eigenvalue of $H$  and the vectors
$|\Phi_k\rangle$ and $\langle \Phi_k^*|$ denote its right and left
eigenfunctions, respectively). They have to be normalized therefore by
means of the complex value  $\langle \Phi_k^*|\Phi_l\rangle$ what is in
contrast to the normalization of the eigenfunctions of a Hermitian
operator by means of the real value $\langle \Phi_k|\Phi_l\rangle$
(for details see sections 2.2 and 2.3 of \cite{top}).
We use the normalization 
\begin{eqnarray}
\langle \Phi_k^*|\Phi_l\rangle = \delta_{kl} 
\label{int3}
\end{eqnarray}
in analogy to $\langle \Phi_k|\Phi_l\rangle = \delta_{kl}$ for
discrete states in order to describe a smooth transition from the
closed system with discrete states and 
orthogonal eigenfunctions to the weakly open system with
narrow resonance states and biorthogonal eigenfunctions. 
The $\Phi_k$ contain (like the $\ce_k$)  global features that are 
caused by many-body forces  induced by the coupling
$\omega_{kl}$ of the two states $k$ and $l\ne k$ via the environment.
Moreover, they contain  the self-energy contributions of the states $k$
due to their coupling $\omega_{kk}$ to the environment. 

The biorthogonality of the eigenfunctions $\Phi_k$ of the non-Hermitian
operator $\ch^{(2)}$ is quantitatively expressed by the ratio 
\begin{eqnarray}
r_k ~\equiv ~\frac{\langle \Phi_k^* | \Phi_k \rangle}{\langle \Phi_k 
| \Phi_k \rangle} \; . 
\label{eif11}
\end{eqnarray}
Usually  $r_k \approx 1$ for decaying states which are well
separated  from other decaying states (according to the
fact that Hermitian quantum physics is a good approach at low level 
density and small coupling strength to the environment). 
Here, the eigenfunctions are (almost) orthogonal,  
$\langle \Phi_k^* | \Phi_k \rangle   \approx
\langle \Phi_k | \Phi_k \rangle   = 1 $.

The situation changes however completely when an EP is approached due
to the fact that the eigenfunctions of two crossing states are linearly
dependent according to (\ref{eif5}) and, by using the normalization condition
(\ref{int3}),
$\langle \Phi_k | \Phi_k \rangle  \to \infty $ at the EP \cite{top}.
Thus, the phases of the two eigenfunctions
relative to one another change dramatically 
when the crossing point is approached and $r_k \to 0$.
The  non-rigidity $r_k$ of the phases of the eigenfunctions of $\ch^{(2)}$ 
follows, eventually, from the fact that $\langle\Phi_k^*|\Phi_k\rangle$
is a complex number (in difference to the norm
$\langle\Phi_k|\Phi_k\rangle$ which is a real number) 
such that the normalization condition
(\ref{int3}) can be fulfilled only by the additional postulation 
Im$\langle\Phi_k^*|\Phi_k\rangle =0$ (what corresponds to a rotation). 

The value  $r_k$, defined by (\ref{eif11}), 
is called {\it phase  rigidity}  of the eigenfunction $\Phi_k$
\cite{top}. Generally $1 ~\ge ~r_k ~\ge ~0 $.  
When $r_k<1$, an analytical expression for the eigenfunctions as 
function of a certain control parameter  can, generally, not be
obtained. The  non-rigidity $r_k<1$ of the phases of the eigenfunctions
of $\ch^{(2)}$ in the neighborhood of EPs is the most important
difference between the  non-Hermitian quantum physics and the Hermitian
one. Mathematically, it causes nonlinear effects in quantum systems 
in a natural manner \cite{top}.
Physically, it gives  one of the  states of the system the possibility 
to align  near to  (and at) the EP with the common environment and to
receive thereby a large width $\Gamma_k$. This alignment is nothing but a
quantitative measure of the influence of the environment onto
the spectroscopic properties of the system \cite{top}. The aligned
state is the short-lived state caused by avoided resonance overlapping.
Its formation is accompanied by trapping the remaining 
resonance states which are long-lived, i.e. decoupled more or 
less from the environment. 

The relations (\ref{eif5}) and $r_k < 1$ in approaching an EP are
seen in experimental results obtained for two resonance states in 
microwave billiards \cite{demb1,demb2}, see \cite{top,fdp1}.
In \cite{berggren}, wave transport in an open non-Hermitian
quantum dot with $\cpt$ symmetry is calculated. The numerical
results show clearly not only the relations (\ref{eif5}) in
approaching an EP  but also the phase rotation, which takes place 
in the vicinity of an EP and is described by $r_k < 1$ as stated above.

In order to receive a statement on the entanglement of the
wavefunctions, it is meaningful to represent
the  eigenfunctions $\Phi_i$ of ${\cal H}^{(2)}$  in the
set of basic wavefunctions $\Phi_k^0$ of ${\cal H}^0$
\begin{eqnarray}
\Phi_k=\sum_{l=1}^N b_{kl} \Phi_l^0 ~~ ;
\quad \quad b_{kl} = |b_{kl}| e^{i\theta_{kl}} \, .
\label{int20}
\end{eqnarray}
Also the $b_{kl}$ are normalized  according to the biorthogonality
relations  of the wavefunctions $\{\Phi_k\}$. The angle $\theta_{kl}$
can be determined from
${\rm tg}(\theta_{kl}) = {\rm Im}(b_{kl}) / {\rm Re}(b_{kl})$ .
The entanglement of the wavefunctions is large in the neighborhood of EPs as
numerical calculations have shown, e.g. \cite{elro2,elro3}.

\section{The $S$-matrix in the vicinity of an exceptional point}
\label{crosssection}

The cross section can be calculated by means of the $S$-matrix 
$\sigma (E) \propto |1-S(E)|^2$.  
In the vicinity of a single level coupled to one channel 
the line shape is  the well-known  Breit-Wigner shape,
\begin{eqnarray}
\label{breitwigner1}
S=1+i\, \frac{\Gamma_k}{E-E_k -\frac{i}{2}\Gamma_k}
\end{eqnarray}
where $E$ is the energy and $E_k$ and $\Gamma_k$ are defined in
  Eq. (\ref{int6}). 
This expression can be rewritten as \cite{ro03}  
\begin{eqnarray}
\label{breitwigner2}
S = \frac{E-E_1+\frac{i}{2}\Gamma_1}{E-E_1-
\frac{i}{2}\Gamma_1}
\end{eqnarray}
which is explicitly unitary.  
Extending the problem to that of two closely neighboring
resonance states that are coupled 
to a common continuum, the representation (\ref{breitwigner2})
of the $S$-matrix reads (up to a background term)
\begin{eqnarray}
\label{smatr}
S = \frac{(E-E_1+\frac{i}{2}\Gamma_1)~(E-E_2+\frac{i}{2}\Gamma_2)}{(E-E_1-
\frac{i}{2}\Gamma_1)~(E-E_2-\frac{i}{2}\Gamma_2)}.
\end{eqnarray}
At an EP,  the $S$-matrix has a double pole. Here (\ref{smatr}) can be
rewritten as \cite{ro03},
\begin{eqnarray}
\label{smatr2}
S = 1+2\,i\,\frac{\Gamma_d}{E-E_d-\frac{i}{2}\Gamma_d}-
\frac{\Gamma_d^2}{(E-E_d-\frac{i}{2}\Gamma_d)^2}
\end{eqnarray}
where $E_1=E_2\equiv E_d$ and $\Gamma_1=\Gamma_2\equiv \Gamma_d$.
The  second term on the right-hand side of this expression corresponds
to the usual linear term (\ref{breitwigner1}) describing a single-state  
Breit-Wigner resonance, however multiplied by a factor
two. The third term is quadratic in  energy. In the cross
section, an interference minimum appears at the EP and the 
two peaks at both sides are asymmetric \cite{mudiisro,elro3}. 
The interference with a direct scattering (background) part may change
the picture when the scattering phase differs from zero \cite{marost03}.

It is interesting to trace the line shape of a resonance  when
the distance to another resonance is varied \cite{elro3}. According to
(\ref{smatr2}) the line shape is of standard  
symmetrical Breit-Wigner form only when both resonances are well
separated from one another. In approaching an EP, the two resonances
avoid overlapping, the $S$-matrix  approaches the expression
(\ref{smatr2}) at the EP and two asymmetric bumps (``resonances'') 
appear in the cross section. Finally, the cross
section shows a broad resonance with a narrow dip in the center 
which is  reminiscent of the long-lived resonance state caused by
width bifurcation together with the short-lived (broad) resonance
state by which it is superposed. 
Thus, the cross section varies smoothly in the whole parameter range
including the critical region around the EP. It is difficult therefore
to trace the influence of an EP on the dynamics of open quantum
systems by considering only the cross section ($S$-matrix). In order 
to receive more information,  the eigenvalues and eigenfunctions of the
non-Hermitian operator  should be
considered which behave non-analytically at the EP (in spite of the
smooth  behavior of the $S$-matrix) and $\langle\Phi_k|\Phi_k\rangle
\to \infty$, respectively.

\section{Special case with imaginary coupling} 
\label{special}

Generally, the expression $Z$ defined in (\ref{int6}) is complex. By 
using the condition $Z=0$, the EP can be found (when it is far from
another EP \cite{top}).  Mostly the critical value $\omega =
\omega^{\rm cr}$ is complex.  In the limiting case with
real $\omega =  \omega_r , ~\gamma_1 = \gamma_2$ and 
$e_1\ne e_2$, no EP exists because $Z^2 > 0$. 

When however $\omega = i \, \omega_i $ is imaginary, 
\begin{eqnarray}
Z = \frac{1}{2} \sqrt{(e_1-e_2)^2 - \frac{1}{4} (\gamma_1-\gamma_2)^2 
+i(e_1-e_2)(\gamma_1-\gamma_2) - 4\omega_i^2} \, ,
\label{int6i}
\end{eqnarray}
the condition $Z=0$ can be fulfilled  
when $(e_1-e_2)^2 - \frac{1}{4} (\gamma_1-\gamma_2)^2 = 4\omega_i^2$
and $(e_1-e_2)(\gamma_1-\gamma_2) =0$, i.e. 
when $\gamma_1 = \gamma_2$ (or when $e_1=e_2$). Let $e_i=e_i(a)$ be
dependent on a certain parameter $a$ and $\omega = i \, \omega_i$ be fixed,
then 
\begin{eqnarray}
(e_1(a) - e_2(a))^2 -4\, \omega_i^2 &= &0 
~~\rightarrow ~~e_1(a) - e_2(a) =\pm \, 2\, \omega_i 
\label{int6b}
\end{eqnarray}
and two EPs appear.  It holds further
\begin{eqnarray}
\label{int6c}
(e_1(a) - e_2(a))^2 >4\, \omega_i^2 &\rightarrow& ~Z ~\in ~\Re \\
\label{int6d}
(e_1(a) - e_2(a))^2 <4\, \omega_i^2 &\rightarrow&  ~Z ~\in ~\Im 
\end{eqnarray}
independent of the parameter dependence of the $e_i$.
In the first case, the eigenvalues ${\cal E}_i = E_i+i/2\, \Gamma_i$ 
differ from the original values 
$\varepsilon_i = e_i + i/2~\gamma_i$ by a contribution to
the energies and in the second case by a contribution to the widths. 
The width bifurcation starts in the very neighborhood of one of the EPs and
becomes maximum in the middle between the two EPs.
This happens at the crossing point $e_1 = e_2$ where 
$\Delta \Gamma/2 \equiv |\Gamma_1/2 - \Gamma_2/2| = 4\, \omega_i$.
A similar situation appears when $\gamma_1 \approx \gamma_2$ as 
results of numerical calculations  show. For details see 
\cite{elro3,acta}.

\section{Inclusion of gain}
\label{gain}

\begin{figure}[ht]
\begin{center}
\includegraphics[width=12cm,height=12cm]{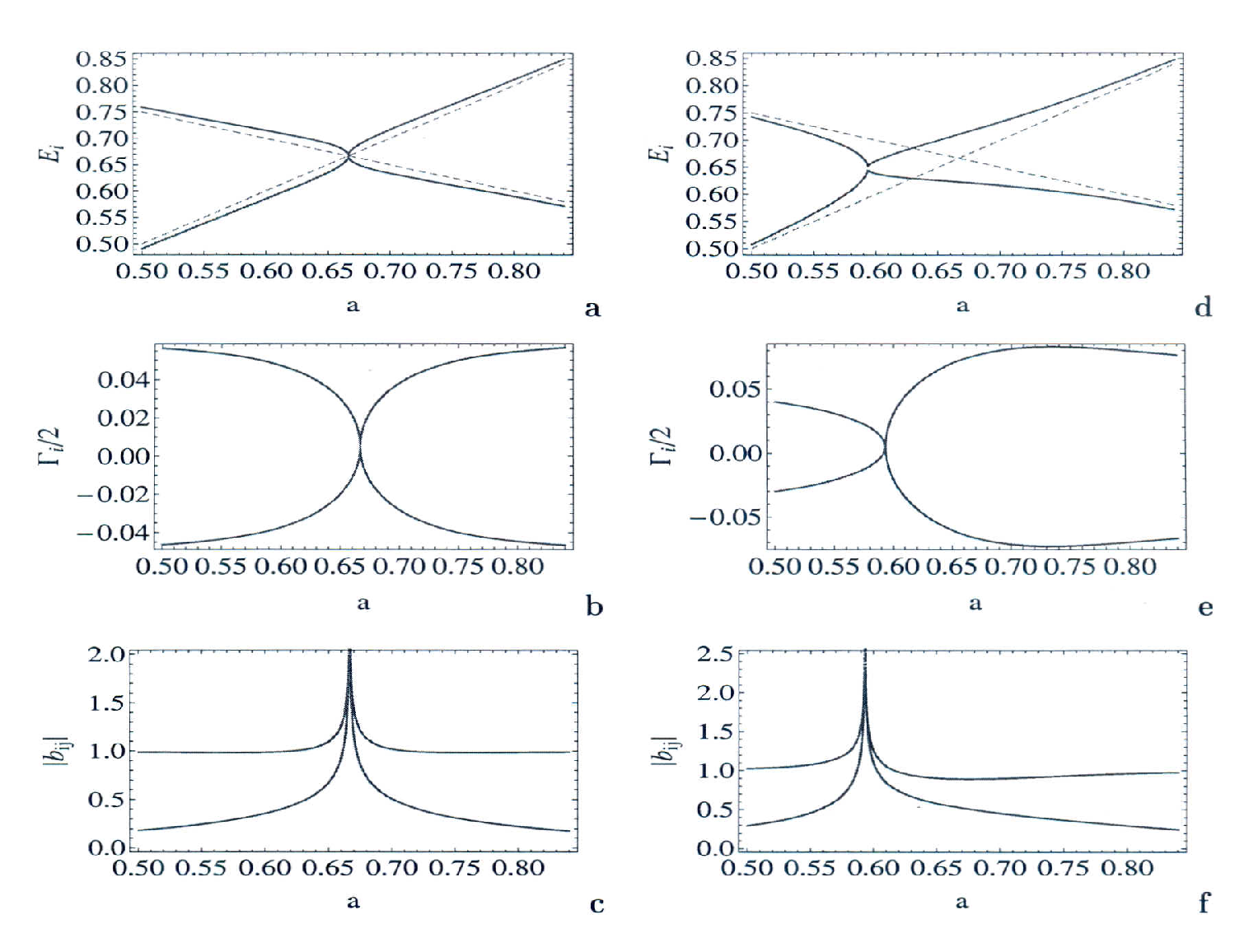}
\end{center}
\caption{\footnotesize
Energies $E_i$ (a, d),  widths $\Gamma_i/2$ (b, e)
and mixing coefficients $|b_{ij}|^2$ (c, f) 
of $N=2$ states coupled to $K=1$ channel as a function of the parameter $a$. 
The parameters are 
$e_{1}=1-a/2, ~e_2= a$; $\gamma_{1}/2=-0.05$; $\gamma_2/2 = 0.06 $; 
$\omega=0.055 $ (left panel); $\omega=0.0789 (1+i)/\sqrt{2} $ (right panel).
The dashed lines show $e_i(a)$.
}
\label{fig1}
\end{figure}

We start with the  symmetric $2\times 2$ matrix (\ref{form1}) 
by assuming a different sign for $\gamma_1$ and $\gamma_2$
according to loss ($\gamma_i \le 0$) and gain ($\gamma_i \ge 0$). 
The two eigenvalues of ${\cal H}^{(2)}$ are given by (\ref{int6}) and 
$Z$ reads
\begin{eqnarray}
Z = \frac{1}{2} \sqrt{(e_1-e_2)^2 - \frac{1}{4} (\gamma_1-\gamma_2)^2 
+i(e_1-e_2)(\gamma_1-\gamma_2) + 4\omega^2} 
\label{int6ig}
\end{eqnarray}
where $\omega = \omega_r + i \omega_i$ is complex, generally. 
As for a natural open system, an EP appears when the
condition $Z= 0$ is fulfilled. However  $\gamma_1 \ne \gamma_2 $ 
and therefore $(\gamma_1-\gamma_2)^2 \ne 0$  
even when $|\gamma_1| = |\gamma_2| $.  The conditions read
\begin{eqnarray}
\label{intgain1}
(e_1-e_2)^2 - \frac{1}{4} (\gamma_1 - \gamma_2)^2& =& - 4 \omega_r^2 + 4
\omega_i^2 \\
\label{intgain2}
(e_1-e_2)(\gamma_1-\gamma_2)&=&-8\omega_r\omega_i
\end{eqnarray}
which  can mostly be fulfilled also when both values 
$\gamma_i$, i.e. loss {\it and} gain, are 
parameter independent and only the energies $e_i$ are parameter dependent
(e.g. $e_i = e_i(a)$ where $a$ is a suitable parameter). Two examples with
real and complex $\omega$, respectively, are shown in
Fig. \ref{fig1}. In both cases, one EP can be seen.

\begin{figure}[ht]
\begin{center}
\includegraphics[width=12cm,height=12cm]{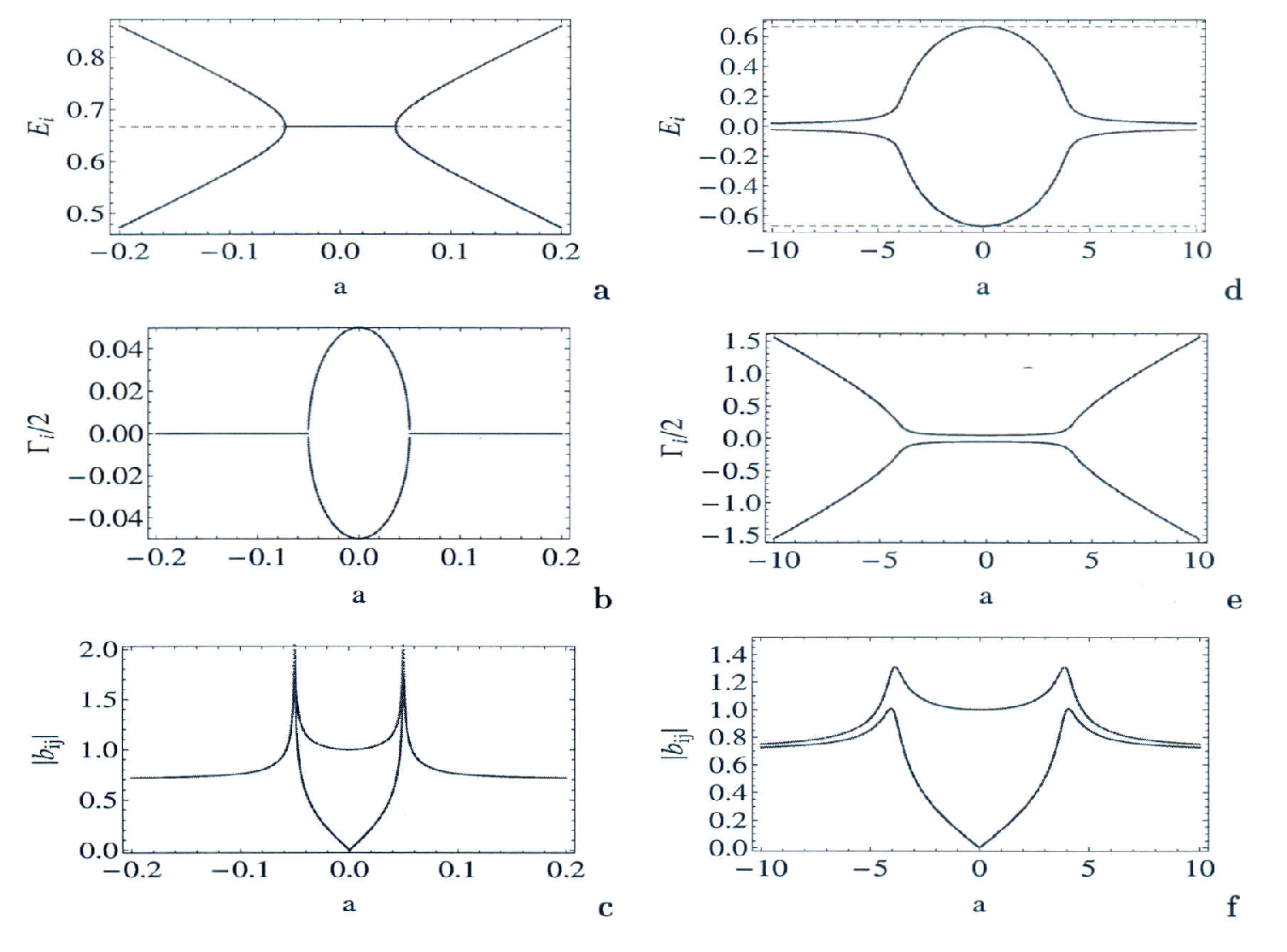}
\end{center}
\caption{\footnotesize
Energies $E_i$ (a, d),  widths $\Gamma_i/2$ (b, e)
and mixing coefficients $|b_{ij}|^2$ (c, f) 
of $N=2$ states coupled to $K=1$ channel as a function of the parameter $a$. 
The parameters are 
$e_{1}=2/3, ~e_2= 2/3$; $\gamma_{1}/2=-0.05$; $\gamma_2/2 = 0.05 $; 
$\omega=a $ (left panel); 
$e_{1}=2/3, ~e_2= -2/3$; $\gamma_{1}/2=-0.05$; $\gamma_2/2 = 0.05 $; 
$\omega=i\cdot a $ (right panel).
}
\label{fig2}
\end{figure}

Let us consider now the special case with parameter dependent coupling 
strengths $\omega = \omega(a)$ and fixed (parameter independent) $e_i$ and
$\gamma_i$. If $e_1=e_2$ and  
$\omega = \omega_r(a)$, it follows from (\ref{int6ig})
\begin{eqnarray}
-\frac{1}{4}(\gamma_1 - \gamma_2)^2 +4\, \omega_r(a)^2 &= &0 
~~\rightarrow ~~\gamma_1 - \gamma_2 =\pm \, 4\, \omega_r(a) 
\label{int6bg}
\end{eqnarray}
and two EPs appear. If $\gamma_1 = - \gamma_2 \equiv \gamma$ the
condition (\ref{int6bg}) reads $\gamma /2 = \pm  \omega_r(a)$. Further 
\begin{eqnarray}
\label{int6cg}
\frac{1}{4}(\gamma_1 - \gamma_2)^2 = \gamma^2 >4\, \omega_r(a)^2 &\rightarrow& 
~Z ~\in ~\Im \\
\label{int6dg}
\frac{1}{4}(\gamma_1 - \gamma_2)^2 = \gamma^2
<4\, \omega_r(a)^2 &\rightarrow&  
~Z ~\in ~\Re 
\end{eqnarray}
independent of any parameter dependence of the $e_i$ and $\gamma_i$. 
In the first case, the eigenvalues ${\cal E}_i = E_i+i/2\, \Gamma_i$ 
differ from the original values 
$\varepsilon_i = e_i + i/2~\gamma_i$ by a contribution to
the widths and in the second case by a contribution to the energies. 
The width bifurcation starts in the very neighborhood of one of the EPs and
becomes maximum in the middle between the two EPs. Numerical results
support this picture (Fig. \ref{fig2} left panel).
According to results of further numerical calculations, a
similar situation appears when $\gamma_1 \approx \gamma_2$ and (or)
$e_1 \approx e_2$.

When $\omega = i\,\omega_i(a)$, the condition $Z=0$ cannot be
fulfilled in the considered special case since 
\begin{eqnarray}
\label{int6cgim}
\frac{1}{4}(\gamma_1 - \gamma_2)^2 +4\, \omega_i(a)^2
= \gamma^2  +4\, \omega_i(a)^2   > 0 
\end{eqnarray}
and no EP exists.
This scenario is analogue to that obtained for a natural open quantum
system when $\omega$ is real, see Sect.  \ref{special}.  This
correspondence is in agreement with Eqs. (\ref{form11}) and (\ref{form12}), 
according to which $\omega$ is complex (almost imaginary) in the first case
and real in the second case. 

As a result, we have two EPs in an open system with two decaying states
($\gamma_1=\gamma_2\ne 0$) when $\omega = i\, \omega_i$ is imaginary. 
In contrast, there is no EP when  $\omega = \omega_r$ is real. 
The situation in an open system with gain and loss
($\gamma_1=-\gamma_2\ne 0$) is the opposite way around\,: there is no
EP when  $\omega = i\, \omega_i$ is imaginary, and two EPs when 
$\omega = \omega_r$ is real. In both cases with two EPs we have width
bifurcation between the two EPs. In the more realistic case with 
complex $\omega$, the results are similar to those discussed above
when  $\omega_i \gg \omega_r$ and $\omega_i \ll \omega_r$, respectively.

Interesting are the results shown in Fig. \ref{fig2} right panel.
Here $e_2 = - e_1$ and $\gamma_2 = - \gamma_1$ are parameter
independent while $\omega(a) = i\, \omega_i(a)$ is imaginary and
parameter dependent. The results are
similar to those for a $\cpt$ symmetric system with parameter dependent
$\gamma_1(a')=-\gamma_2(a')$, fixed $e_1=e_2$ and fixed real  $\omega$
(see Fig. 1 left panel in \cite{acta} where $\Gamma_i = 0$ in the finite
parameter range between the two EPs according to $\cpt$ symmetry and its
breaking at the EPs). In order to
see the EPs and $\Gamma_i = 0 $ in the parameter range between them 
also in the present case, $\omega$ has however to be complex
because (\ref{int6cgim}) holds true when $\omega = i\,\omega_i$. The 
two critical values $\omega_i(a^{\rm cr})$ and $\omega_r(a^{\rm cr})$ at the EPs
can be determined by solving the two equations 
(\ref{intgain1})   and (\ref{intgain2}) with the fixed values $e_i$
and $\gamma_i$. The numerical results support the analytical ones.
The results are however  sensitive relative to small variations of 
the parameter $a$. In this manner, Fig. \ref{fig2} illustrates  that different 
situations with  gain  and loss
can be realized in open quantum systems. A few of them
are sensitive relative to small variations of $a$
(e.g. Fig. \ref{fig2} right panel) while most of them are more stable
(e.g. Fig. \ref{fig2} left panel).

It should be underlined here the following.
In difference to a $\cpt$ symmetric system (see Fig. 1 left panel 
in \cite{acta})
the relations (\ref{int6bg}) to (\ref{int6dg}) are obtained
by keeping fixed the widths $\gamma_i$ (and the energies $e_i$),
and varying the coupling strengths $\omega = \omega(a)$
by means of the parameter $a$. This condition is, probably, easier to
realize experimentally than the parametrical dependence 
$\gamma_i(a')$ of the widths. Remarkably is the possibility 
to realize very different situations in an open quantum
system (compare left and right panels of Fig. \ref{fig2}).

In any case, the mathematical condition for an EP to occur  
in an open quantum system with loss and gain, is the same as that for 
a natural open quantum system, namely  $Z=0$. The only difference is
that the two widths $\gamma_i$ have different sign in the first case
while they have the same sign in the last case. In both cases, the EP
influences a finite parameter range in its neighborhood.

\section{Conclusions}
\label{concl}

In the present paper we have considered generic properties of 
open quantum systems which are embedded, in a natural manner, in the 
continuum of decay channels. Due to the coupling of the system states to
the environment, the Hamiltonian $\ch$ describing the  system is  
non-Hermitian. The eigenvalues $\ce_k$ of $\ch$  are complex, generally,
and provide not only the energies Re$(\ce_k)$ of the
states but also their lifetimes being inverse proportional to the 
widths Im$(\ce_k)$. Although the eigenfunctions $\Phi_k$ of $\ch$ are
biorthogonal, they can be normalized by $\langle
\Phi_k^*|\Phi_l\rangle = \delta_{kl}$ such that the transition from a
weakly opened system with narrow resonance states to a strongly opened
system with (at least) one short-lived resonance state occurs
smoothly. This transition is possible because the phases 
of the eigenfunctions of $\ch$ are not rigid\,: while 
the wavefunctions of two states are (almost)
orthogonal to one another at large distance from the EP (as those of the
corresponding closed system), they become linearly dependent from one
another in approaching the EP. In the vicinity of an EP, the
environment is able to put information into the system by means of 
aligning one of the system states to
the states of the environment and, at the same time, decoupling the
remaining states from the environment. That means, a short-lived
(the {\it aligned}) state appears together with long-lived 
(the decoupled, {\it trapped}) states, a process called mostly 
width bifurcation. The changes of the phase rigidity and of the 
lifetime of the states near to (or at) an EP 
are proven by Figs. 4 and 5 of \cite{berggren} obtained 
for an open  quantum billiard with $\cpt$ symmetry.
Furthermore, the eigenfunctions of $\ch$ are
strongly mixed (entangled) in a finite parameter range around an EP 
such that a nearby state will have a large influence not only
in this parameter range but also somewhat beyond this range. A dynamical phase
transition may therefore occur in the system which changes radically
the spectroscopic properties of the system
(a detailed discussion of the relation between EPs and dynamical
  phase transitions can be found in \cite{elro4}). The dynamical phase 
transition has the same characteristic features which are known from  
$\cpt$-symmetry breaking occurring in a $\cpt$-symmetric system under
some critical condition. 

In the second part of the paper, we have studied the properties of open
quantum systems by including ``gain'' from the environment. This 
is, of course, not realized in nature. As the results obtained 
some years ago for $\cpt$-symmetric optical lattices 
\cite{mumbai1,mumbai2} 
and recent results for other systems show, it can however be realized 
in experiment, e.g.  \cite{schindler,bender}. The results of our
theoretical studies presented in the present paper, indicate
that the consideration of gain is not restricted to  $\cpt$-symmetric
systems. It is a much more general phenomenon appearing in open quantum
systems that are described by a non-Hermitian Hamiltonian.
The conditions for desired properties can be formulated. 
In particular, it is possible to consider different layers with
fixed (parameter independent) loss or gain which is characteristic of
the layer in question. These results are of importance
for basic research as well as for applications.

\end{document}